\documentclass[12pt]{article}
\usepackage{natbib}
\pdfoutput=1 
\usepackage{amsmath,amsfonts,amssymb}
\usepackage[english]{babel}
\usepackage{graphicx}
\usepackage{color}
\newcommand{\ra}{\rightarrow}

\newcommand{\lra}{\longrightarrow}
\newcommand{\Lra}{\Longrightarrow}
\newcommand{\half}{{\textstyle{\frac{1}{2}}}}
\newcommand{\third}{{\textstyle{\frac{3}{2}}}}
\newcommand{\five}{{\textstyle{\frac{5}{2}}}}
\newcommand{\seven}{{\textstyle{\frac{7}{2}}}}

\newcommand{\JN}{J^{N}}
\newcommand{\wh}{\widehat}

\newcommand{\be}{\begin{equation}}
\newcommand{\ee}{\end{equation}}
\newcommand{\bea}{\begin{eqnarray}}
\newcommand{\eea}{\end{eqnarray}}

\begin{document}

\bibliographystyle{plainnat}


\pagestyle{empty}

\bigskip

\bigskip

\begin{center}

\vfill 
 
 \vspace{30mm}

{\Large \textbf{\textsf{Metrics and Harmonic Analysis on DNA}}}

\vspace{10mm}

{\large   A. Sciarrino}

\vspace{10mm}

 \emph{I.N.F.N., Sezione di Napol  \\ Complesso Universitario di Monte S. Angelo \\ Via Cinthia, I-80126 Napoli, Italy \\ nino.sciarrino@gmail.com}

\vspace{12mm}

\end{center}

\vspace{7mm}

\begin{abstract}
 A mathematical algorithm to describe DNA or RNA  sequences of $N$ nucleotides  by a string of $2N$ integers numbers is presented in the framework of the so called crystal basis model of the genetic code. The description allows  to define  a not trivial distance between two sequences and to perform  the Discrete Fourier Transform on a sequence.  Using the definition of distance the sequence of $\beta$-globin for Homo, Gallus and Opossum are compared. The Discrete Fourier Transform for  some simple examples  is  computed.
\end{abstract}

\vfill
{\bf Keywords}: 
DNA, RNA, nucleotide sequences,  crystal basis model, genetic code, distance,   discrete Fourier transform  


\newpage

\pagestyle{plain}
\setcounter{page}{1}

\section{Introduction}

\hspace{0.5 cm}  In the last years, complete genomes have been identified and the expressions of tens of thousands of genes for many  organisms  have been identified.
 
As more genomes become sequenced and more genes are located, the  identification and characterization of gene functions  are the great new challenges of the functional genomics. 
 
The advent of large-scale ``transcriptomic" and ``proteomic"  technologies  is providing each day huge amounts of data, pushing to search new approaches to  discover and characterize gene expression patterns and, more generally,  nucleotide sequences in DNA as well as in RNA. Moreover, with the advent of computers,  new scientific branches have arisen, the bioinformatics and next the bioengineering. As emblematic example, let us quote the MIT programming language  to design DNA circuit  \cite{MIT},  which very  recently has gathered a very wide appeal in  the world media  for its implication on the genetic manipulations.  So it appears evident the need to translate any DNA or RNA sequence in a mathematical language,  in order to allow its study  by means of the tools of the mathematics and/or of the artificial intelligence.

The standard representation of nucleotide sequences is by letter-series using the   four letter ``alphabet''  \{C,T/U,G,A\} of the DNA/RNA. However,  the availability of  DNA and RNA data, each day extremely fast  growing, has prompted in the last decades, a research toward mathematical and graphical representation of nucleotide series in order to quickly extract  relevant information. At our knowledge one of the first alternative representation has been proposed in  \cite{HR}, where a method has been described  through which any sequence is represented by a three dimensional curve. In \cite{NHB} several graphical and numerical methods are reviewed and compared.

Another very relevant question arising when nucleotide sequences are compared is the characterization of their differences, in other words their  ``distance''.
 
The first and commonly used measure between two sequences is the Hamming distance \cite{Hamming}. The Hamming distance between two sequences of{ \bf equal length} is the number of positions at which the corresponding symbols are different. In other words it measures the minimum number of substitutions required to change one sequence into the other. The most simple generalization of the Hamming distance is the Kimura 2-parameter distance \cite{Kimura80} assigning different parameter if a pyrimidine Y  (Y=C,T/U) (a purine R (R=A,G)) nucleotide is replaced by the other one, transition, or by a purine R, transversion,  (resp. pyrimidine Y).  

Given two symbolic  sequences  of {\bf different length}, their relative distance, that is a measure of the amount of difference between them, is commonly computed by the Levenshtein distance  \cite{Leve65}, which is defined as the minimum number of  ``edit operation'' needed to transform one sequence into the other,  the  ``edit operation'' being insertion, deletion, or substitution of a single symbol.\footnote{The Levenshtein distance between two strings of the same length is strictly less than the Hamming distance.} 

As the measure of the distance between two sequences is extremely relevant in the theory of the evolution, many and more sophisticated measures have been introduced, whose discussion is beyond the aim of the present paper. However these measures, some of which are not strictly measures, depend only on the positions or, at most, by some global property, e.g. the GC content.\footnote {In \cite{SH}  several distance measures  for sequences have been examined and compared.}

In the present work  we introduce  a {\bf  parametrization}  to describe  a DNA or RNA sequence, which provides information on the distribution of the nucleotides in the sequence.
The mathematical scheme establishes a correspondence between any sequence of N nucleotides  and a set of 2N half-integer and integer numbers\footnote{Clearly one can deal only with integer numbers by multiplying  our parameters by 2.}, so DNA/RNA can be considered as a discrete function with codomain in the set of integer  number.

Let us point out some features of our parametrization.
\begin{enumerate}
\item it allows to introduce  a well defined distance between any two sequences;
\item it allows to perform a Discrete Fourier Transform (DFT)  and, consequently, to exploit all the power of the harmonic analysis; 
\item a transition, a  transversions and, more generally, any mutation can be defined as tensor operator, so allowing to have a mathematical description where the mutation of a nucleotide is position dependent.
\end{enumerate}

The aim of the paper is not to discuss specific problems, but to list, surely in a not exhaustive way, the possibilities that our mathematical description provides.
The paper is organized as follows: in Sect. 2 we  concisely recall the mathematical scheme which  is at the base of the parametrization; in Sect. 3 we describe the algorithm which allows to represent the nucleotide sequence in terms of  a discrete function taking value in the set  of half-integer and integer numbers, discussing the main features of the parametrization, presenting also a few simple examples, and in Sect. 4 we give a short summary and  some highlights on future developments, in particular on the above quoted 3-rd point.

\section{Mathematical parametrization of DNA/RNA sequences}

\subsection{Basic mathematical structure of the crystal basis model }

\hspace{0.5 cm} In order to label  a sequence of N nucleotides with 2N half-integer and integer numbers we have to briefly recall the model of genetic code introduced in \cite{FSS1} and called ``crystal basis model''. In the model the 4 nucleotides,   denoted by the four letter set \{C,T/U,G,A\} in DNA/RNA,
  are assigned   to the 4-dim fundamental  irreducible 
 representation (irrep.) of $\mathcal{U}_{q \to 0}(sl(2) \oplus sl(2))$,  the two $sl(2)$ being denoted
   as  $sl_{H}(2) $ and $sl_{V}(2)$.\footnote{The special basis of  $\mathcal{U}_{q }(sl(n)$ in the limit $q \to 0$ is called a {\bf crystal basis}, see \cite{Kashi}.  This explains the name of the model.} The fundamental  irrep., of dimension 2,  is labelled by $(J_H =1/2, J_V=1/2)$  and the states are identified by  the following assignment for the value of  the third component of the operator 
  $J_H$ ($J_V$)  denoted by $J_{3,H} = \pm \half$ (resp. by $J_{3,V} = \pm \half$):
\be 
	\mbox{C} \equiv (+\half,+\half) \qquad \mbox{T/U} \equiv (-\half,+\half) 
	\qquad \mbox{G} \equiv (+\half,-\half) \qquad \mbox{A} \equiv 
	(-\half,-\half)
	\label{eq:gc1}
\ee 
  In \cite{Kashi} it has been shown that the tensor product of two crystal bases labelled by $J_{1}$ and 
$J_{2}$ can be decomposed into a direct sum of crystal bases labelled, as 
in the case of the tensor product of two  standard $sl(2)$ irreducible representations, by a set of integer ($ J_{1} + J_{2}$ integer) or  half-integer   ($ J_{1} + J_{2}$ half-integer)
numbers $J$   such that
\begin{equation}
\vert J_{1} - J_{2} \vert \le J \le J_{1} + J_{2}
\end{equation}
The new peculiar and crucial feature is that, in the limit $ q \to 0$, the basis vectors of the
 $J$-space are \emph{pure states}, that is they are the product of a state belonging to 
the $J_{1}$-space and of a state belonging to the $J_{2}$-space, while in 
the case of $sl(2)$ or of ${\cal U}_{q}(sl(2))$ they are linear 
combinations of states belonging to the $J_{1}$-space and to the $J_{2}$-space with coefficients called respectively Clebsch-Gordan 
coefficients and $q$-Clebsch-Gordan coefficients. It should be noted that in the case of  $\mathcal{U}_{q}(sl(2))$, then in particular for   $\mathcal{U}_{q \to 0}(sl(2))$,  the tensor product is not symmetric and one has to specify  which is the first irrep. in the tensor product, i.e. an order has to be fixed. In the framework of  DNA/RNA the order appears in a very natural way as  the reading frame moves toward
  the $ 5 \prime \ra 3 \prime $ direction.
 
It follows that  a state of an  irrep. contained in the tensor product of $N$ fundamental irreps. is an N-ordered sequence of states belonging to the fundamental irreps., i.e. an ordered sequence of N letters belonging to the  four letter ``alphabet''  \{C,T/U,G,A\},  and is labelled by  $J_a$ and $J_{3,a}$, $a = H, V$:
\bea
&& J_a = J_{(a,Max)},  J_{(a,Max)} - 1, J_{(a,Max)} -2, \ldots  \geq 0 \;\;\;\;\;  J_{(a,Max)} = N/2 \\
&&   J_{3,a} = -J_a, -J_a + 1, \ldots,  J_a
\eea
As for $N > 2$ an  irrep. labelled by $J_a \neq  J_{(a,Max)}$ is degenerate, i.e. there are several irreps. labelled by the same values of $J_a$, we need to identify an irrep. by other $2(N -2)$ labels, which we choose to be the labels of the  tensor product of $k$ ($k = N-1, N-2,  \ldots, 2$) fundamental irreps., obtained by removing $(N - k)$  irreps.  in the tensor product, starting from the right. In the following we denote the J labels by un upper index $k$.

Let us point out that the above given prescription to remove $(N - k)$  irreps. starting from the right is just a convention. One may equally well choose  to remove $(N - k)$  irreps. starting from the left. However the results of the two prescriptions are not, in general, equal and the choice of the prescription modifies, e.g., the distance  between sequences, see next Section. Our chosen prescription is suggested by the direction of the transcription process from DNA to RNA.
 
 Summarizing, making use of the
properties of  {\bf crystal basis},  any string of $N$ binary label (spin)  $ \; x \in \{\pm \equiv R,Y \;  \mbox{or}  \; \equiv S,W \; \mbox{or}  \;   \equiv  C,U   \; \mbox{or}  \;   \equiv  G, A   \;  \mbox{or}  \;   \equiv  C,G   \;  \mbox{or}   \;   \equiv  U,A \} \; $ can be seen as a state of an  
irrep. contained in $N$-fold tensor product of the the 2-dim fundamental 
irrep. (labelled by  $J = 1/2$) of $\; sl_{q  \to 0}(2) \;$  whose state are 
labelled by $ \; J_{3} =  \pm 1/2 \equiv \pm \,$. In the case of the nucleotides,  due to the presence of  two $sl(2)$, eq.(\ref{eq:gc1}), the number of needed labels has to be doubled. Therefore, in the 
most general case, a sequence of N nucleotides  can be identified by the following $2N$ labels. 
\be
{\mbox\{N-nucleotides\}} \Lra 	\mid \mathbf{J}_{H} \mathbf{J}_{V} \rangle = \mid J_{3,H}, J_{3,V}; J_{H} ^N, J_{V} ^N; \ldots ;J_{H} ^2, J_{V} ^2 \rangle
\label{eq:Nnucl}
\ee
where $\JN_{a}$ ($a = H, V$) labels the irrep. of of $\mathcal{U}_{q \to 0}(sl_{a}(2))$ which the state belongs to, $J_{3,a}$ is the value of the 3rd  component of the operator $J_{a} ^N$. Note that  $J^k$ and  $J_3$ are integer for even  $k$  and half-integer for odd  $k$.
 Obviously a H spin flip, i.e. the change by $\pm 1$ of $J_H$,  (V  spin flip) corresponds, respectively,  to a biological transition (transversion).   
 
For example, the codon CGA is represented as
\be
\mid  CGA \rangle = \mid J_{3,H}, J_{3,V}; J_{H} ^3, J_{V} ^3; J_{H} ^2, J_{V} ^2 \rangle =  \; \mid\half , - \half ; \half, \half;   1,  1 \rangle 
\ee
The labels have been computed considering the codon as a composite state resulting from the 3-fold tensor product of the fundamental irrep., tensor product  computed  following the below indicated order
\be
\Bigl(\frac{1}{2},\frac{1}{2} \Bigl) \ \otimes \ \Bigl(\frac{1}{2},\frac{1}{2} \Bigl) \ \otimes  \ \Bigl(\frac{1}{2},\frac{1}{2} \Bigl)  = \Bigl(\frac{1}{2},\frac{1}{2} \Bigl)  \otimes   \Biggl[
\Bigl(\frac{1}{2},\frac{1}{2} \Bigl) \ \otimes \ \Bigl(\frac{1}{2},\frac{1}{2} \Bigl) \Biggl]
\ee
 
 \subsection{Labelling of  an ordered sequence of N nucleotides}

\hspace{0.5 cm} In the following we give the algorithm for the computation of the labels $J_{a}^k$ appearing in eq.(\ref{eq:Nnucl})  for a sequence of $N$ nucleotides.

Denoting by $N_{X}$ the number of nucleotides of type $X \in \{C,U,G,A\}$\footnote{Clearly for DNA U should be replaced by T.}, we have:
\be
2 \, J_{3,H} =  N_{C} + N_{G} - N_{U} - N_{A} \equiv N_{CG} - N_{UA} \;\;\;\;\;\;\; 
2 \, J_{3,V} =  N_{C} + N_{U} - N_{G} - N_{A} \equiv N_{CU} - N_{GA}
\ee
   The  {\bf GC content}, i.e. the number $N_{GC} =  N_{C} + N_{G} $ of G and C nucleotides is  clearly related to the value of 
$J_{3,H}$  by 
\be
J_{3,H} =  N_{GC} \, - \, \frac{N}{2}
\label{eq:J3H}
\ee
while the {\bf asymmetry}, i.e. the difference of the number of the complementary bases or the excess  of purines, is related to the value of $J_{3,V}$  \be
J_{3,V} =  N_{CU} \, - \, \frac{N}{2}
\label{eq:J3V}
\ee
 Let us recall that we have $ |J_{3,H}|  \leq J_{H} ^N \leq N/2$ and $ |J_{3,V}|  \leq J_{V} ^N \leq N/2$, where the equality for H (V) labels holds for a sequence of only  C and G or U and A (resp. C and U or G and A) nucleotides. 

 \begin{itemize} 
\item The {\bf algorithm} - From eqs.(\ref{eq:J3H})-(\ref{eq:J3V}) we can compute the values of $J_{3,H}$ and   $J_{3,V}$  corresponding to a fixed sequence of N nucleotides. In the following we describe, in a very didactic way,  the algorithm  to compute the values of  $J_{H} ^N, J_{V} ^N; \ldots ;J_{H} ^2, J_{V} ^2$  corresponding to the sequence of N nucleotides.
The value of  $\JN_H$  is computed replacing, in the sequence, the nucleotides C, G by the symbol $S$ and A, U/T by the symbol  W (the standard symbols denoting the two couples of nucleotides with respectively  Strong (Weak) hydrogen bond),  then taking away  the  $W$ elements, which are at right of the $S$ elements,
 contracting  each of them with  a $S$ element on the left, and then summing the numbers of left 
  $S$  and  $W$,  which of course are, respectively,  at  the right of $W$ and at the left of $S$.
In other words this value is computed deleting all the couples $SW$ of first neighbours which are present in the
sequence and iterating the procedure, on the  sequence obtained by deleting the first couples $SW$, as long as no $W$ appears on the right of 
 any $S$. We refer to the elements which are deleted in the procedure as \emph{contracted}. 
 The value $\JN_V$ is  computed replacing in the sequence the nucleotide C,  U/T by $Y$ and A,G by  $R$  (the standard symbols for pyrimidines and purines)  and then proceeding in the same way as  for the computation of  $\JN_H$, but now deleting all the couples  $YR$, that is $R$ ($Y$) takes the role of $W$ (resp. $S$). 
 Then we proceed to compute the value of $J_{H} ^k$ and of $J_{V} ^k$ ($k = N-1, \ldots,2$) deleting in the original sequence $N - k$ nucleotides from the right and applying to the obtained $k$-nucleotide sequence the above given prescriptions.
  The algorithm, nevertheless its apparent complexity, can be easily implemented in computers and is in detail illustrated in the example below.
 \end{itemize}
 
{\bf Example:}  Let us consider the sequence AGGACTTA
 \begin{enumerate}
\item Calculation of $J_{3,H}$ and  $J_{3,V}$.  From eqs.(\ref{eq:J3H})-(\ref{eq:J3V}), we find 
\be
J_{3,H} \equiv N_{CG}  - \frac{N}{2} = 3 \, - \, 4 = -1 \;\;\;\;\, \;\;\;\;\,  J_{3,V}   \equiv N_{CT}  - \frac{N}{2} = 3 \, - \, 4 = -1
 \ee
 \item  Calculation of $J_{H}^k$, ($k = 8, 7, \ldots, 2$). We replace a contracted couple $SW$ by $ \cdot$.
 \bea
 && AGGACTTA \lra WSSWSWWW  \lra WS \cdot  \cdot  \, WW  \lra W  \cdot  \cdot  \cdot  W  \Lra J_H^{8} = 1 \nonumber \\
 && AGGACTT \lra WSSWSWW  \lra WS \cdot  \cdot  \, W  \lra W  \cdot  \cdot  \cdot   \Lra J_H^{7} = 1/2 \nonumber \\
&& AGGACT \lra WSSWSW  \lra WS \cdot  \cdot      \Lra J_H^{6} = 1 \nonumber \\
&& AGGAC \lra WSSWS  \lra WS \cdot S      \Lra J_H^{5} = 3/2 \nonumber \\
&& AGGA \lra WSSW  \lra WS \cdot       \Lra J_H^{4} = 1 \nonumber \\
&& AGG \lra WSS       \Lra J_H^{3} = 3/2 \;\;\;\;AG \lra WS       \Lra J_H^{2} = 1 \nonumber \nonumber \\
 \eea
 \item  Calculation of $J_{V}^k$, ($k = 8, 7, \ldots, 2$). We replace a contracted couple $YR$ by $ \cdot$.
  \bea
 &&  AGGACTTA \ \lra  RRRRYYYR \lra   RRRRYY \cdot  \Lra J_V^{8} = 3 \nonumber \\
 &&  AGGACTT \ \lra  RRRRYYY \lra   RRRRYYY  \Lra J_V^{7} = 7/2 \nonumber \\
 &&   \Lra J_V^{6} = 3   \Lra J_V^{5} = 5/2  \Lra J_V^{4} = 2  \Lra J_V^{3} = 3/2  \Lra J_V^{2} = 1  
 \eea
  \end{enumerate}
Summarizing, the sequence is labelled, the order  of the labels being specified in eq.(\ref{eq:Nnucl}),  by: 
\be
\mid AGGACTTA \rangle = \mid-1,-1;1,3; \half, \seven; 1,3; \third, \five ; 1,2; \third, \third; 1,1 \rangle
\label{eq:8seq}
\ee

{\bf Remarks}:
\begin{itemize}
\item $J_{a}^n$ differs by $J_{a}^{n \pm 1} $ by $\pm \half$, ($n= 2, \ldots, N-1$. $a = H,V$);
 
\item if  $J_{a}^n = J_{a}^{n +  2k}$  ($n \geq 2, k = 1,2, \ldots$) the numbers of  S and W  nucleotides, between the n-th and (n+2k)th-position, are equal and  the W nucleotides are arranged on the left of the S nucleotides (a = H) or  the numbers of  Y and R  nucleotides are equal and the R nucleotides are arranged on the left of the Y nucleotides (a = V);
 
 \item the values of the labels take somehow into account the arrangement of the nucleotides. To further illustrate this point we  report in Table \ref{table-1} the value computed with a string of 5 pyrimidines formed by 2 U and 3 C, in different position. For all the strings the values of the V-labels are: $J_{3,V} = 5/2, J_V^{5} = 5/2, J_V^{4} = 2, J_V^{3} = 3/2, J_V^{2} = 1$.
 
   \begin{table}[htbp]
\begin{center}
\begin{tabular}{|c||c|c|c|c|c|}
\hline
 Sequence & $J_{3,H}$ & $J_H^5$ &    $J_H^4$  &   $J_H^3$ &   $J_H^2$   \\
 \hline\hline
UUCCC & 1/2 & 5/2 & 2 & 3/2 &  1 \\
\hline\hline
CUUCC & 1/2 & 3/2 & 1 & 1/2 &  0 \\
 \hline\hline
 CCUCU & 1/2 & 1/2 & 1 & 1/2 &  1 \\
 \hline\hline
 CCCUU & 1/2 & 1/2 & 1 & 3/2 &  1 \\
  \hline\hline
 CCUUC & 1/2 & 1/2 & 0 & 1/2 &  1 \\
  \hline\hline
 CUCUC & 1/2 & 1/2 & 0 & 1/2 &  0 \\
 \hline
\end{tabular}
\caption{Examples of labeling of sequences with the same content, but different position, of the 5 nucleotides.}
\label{table-1}
\end{center}
\end{table}
\end{itemize}

\begin{figure}[htbp]
\includegraphics[width=15cm]{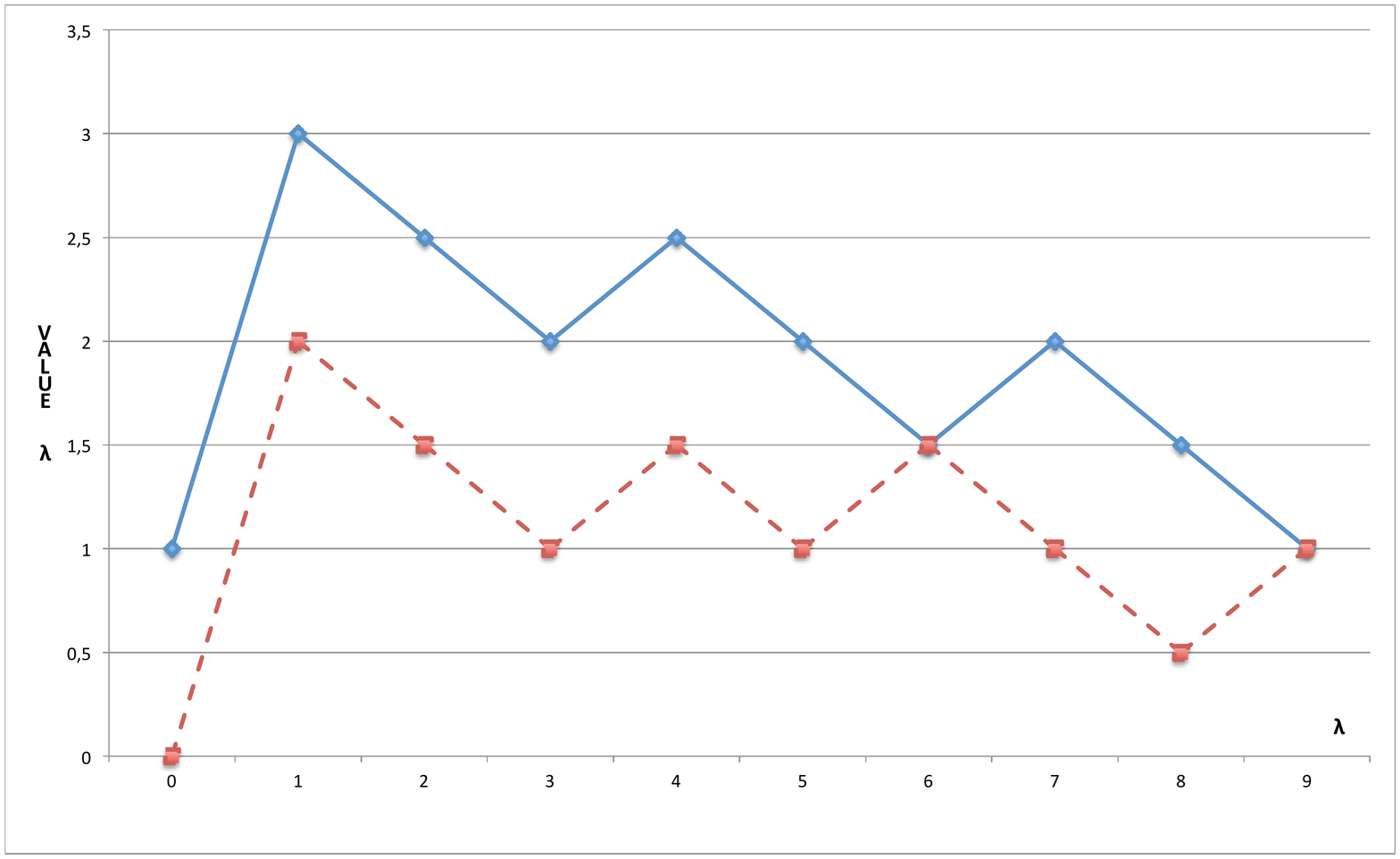}
\centering\caption{Values of  labels $\lambda_{k,H}$, blue continuous lines, and $\lambda_{k,V}$, orange dashed lines, ($k=0,1,\ldots,9$) for the 10 
nucleotides DNA sequence $ATGGTGCACC$.}
 \label{HVlabels}
 \includegraphics[width=15cm]{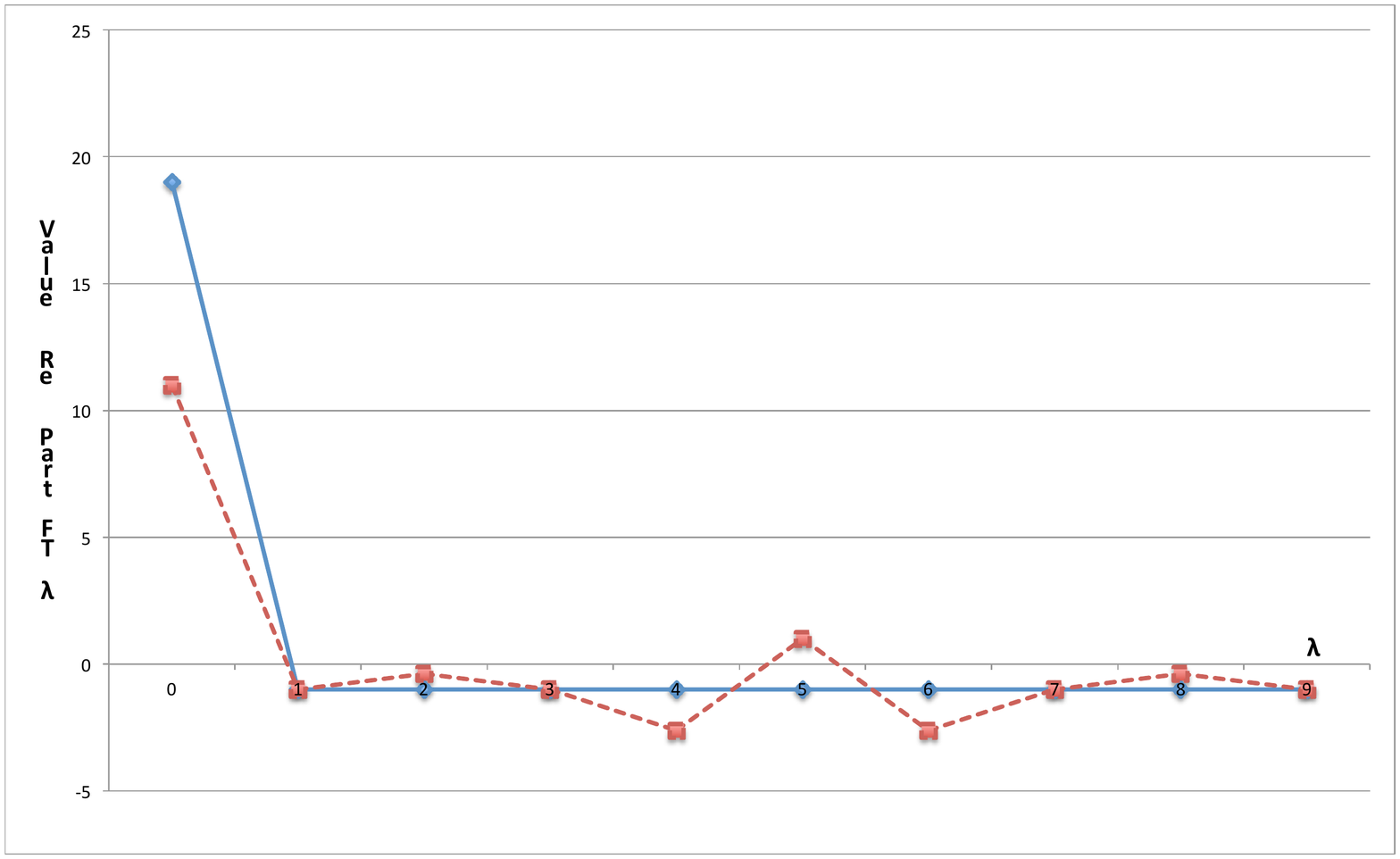}
\centering\caption{Values of the real part of the DFT of the labels $\lambda_{H}^k$, blue continuous lines, and $\lambda_{V}^k$,  orange dashed lines, ($k=0,1,\ldots,9$) for the 10 
nucleotides DNA sequence $ATGGTGCACC$.}
 \label{ReTF}
 \end{figure}

\begin{figure}[htbp]
 \includegraphics[width=15cm]{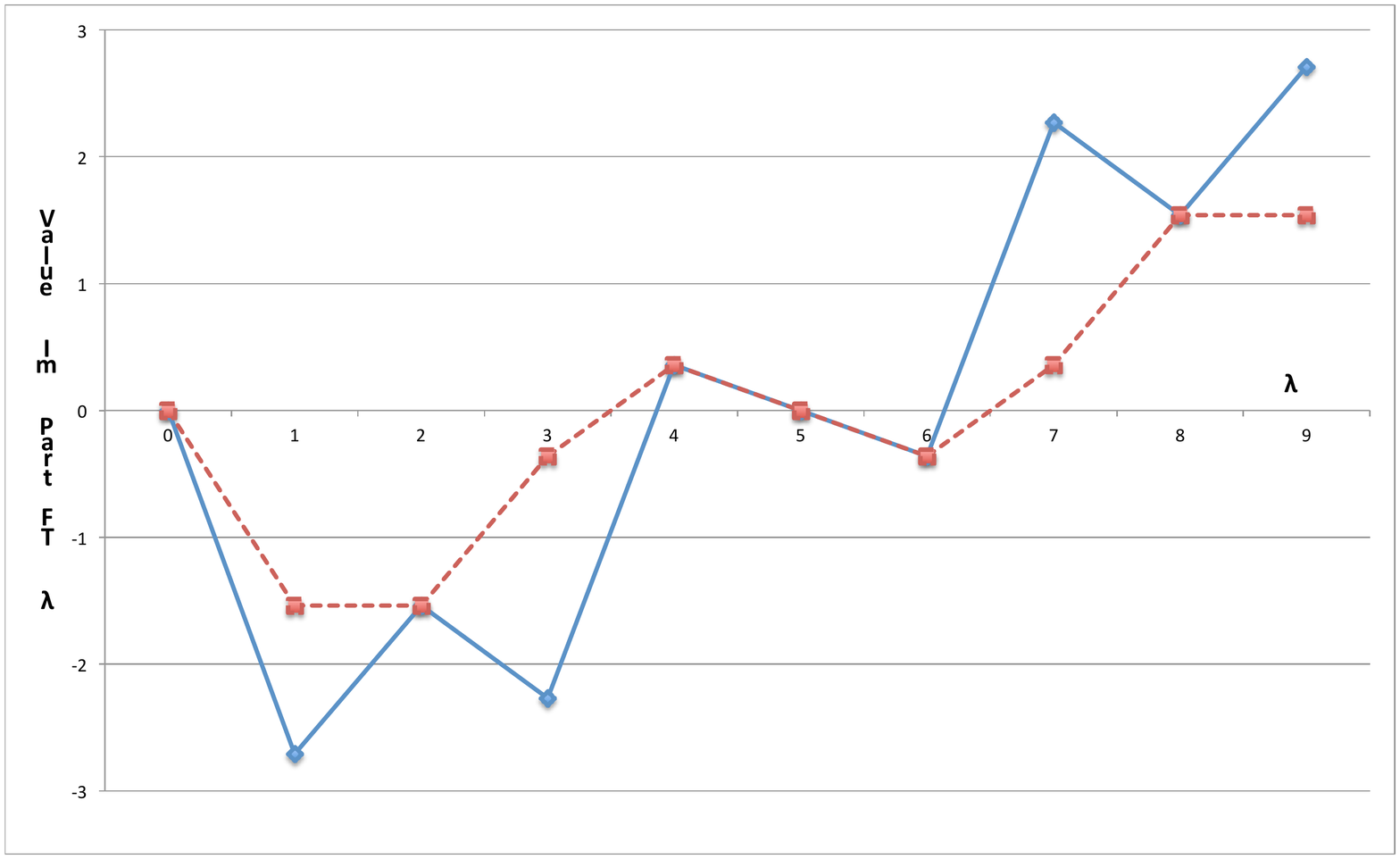}
\centering\caption{Values of the imaginary part of the DFT of the labels $\lambda_{H}^k$, blue continuous lines, and $\lambda_{V}^k$,  orange dashed lines, ($k=0,1,\ldots,9$) for the 10 
nucleotides DNA sequence $ATGGTGCACC$.}
 \label{ImTF}
 \includegraphics[width=15cm]{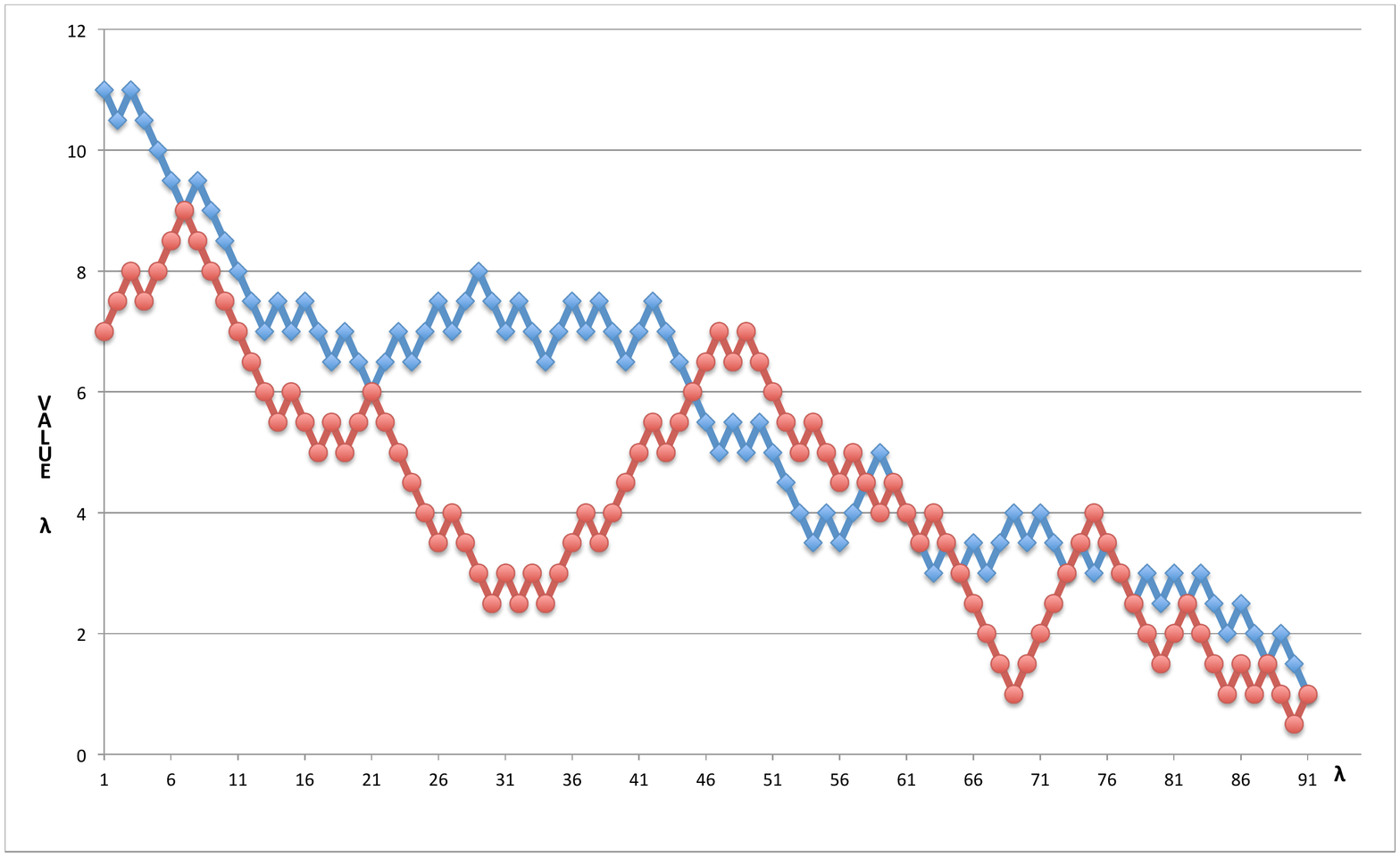}
\centering\caption{Values of labels $\lambda_{k,H}$, blue diamonds, and $\lambda_{k,V}$, orange  circles, ($k=1,\ldots,91$) for the $\beta-globin$ gene for Homo sapiens. The value of  $\lambda_{0,H} =  9$ and $\lambda_{0,V} = -6$  are not reported in the figure.}
 \label{HVbetaH}
 \end{figure}

\begin{figure}[htbp]
 \includegraphics[width=15cm]{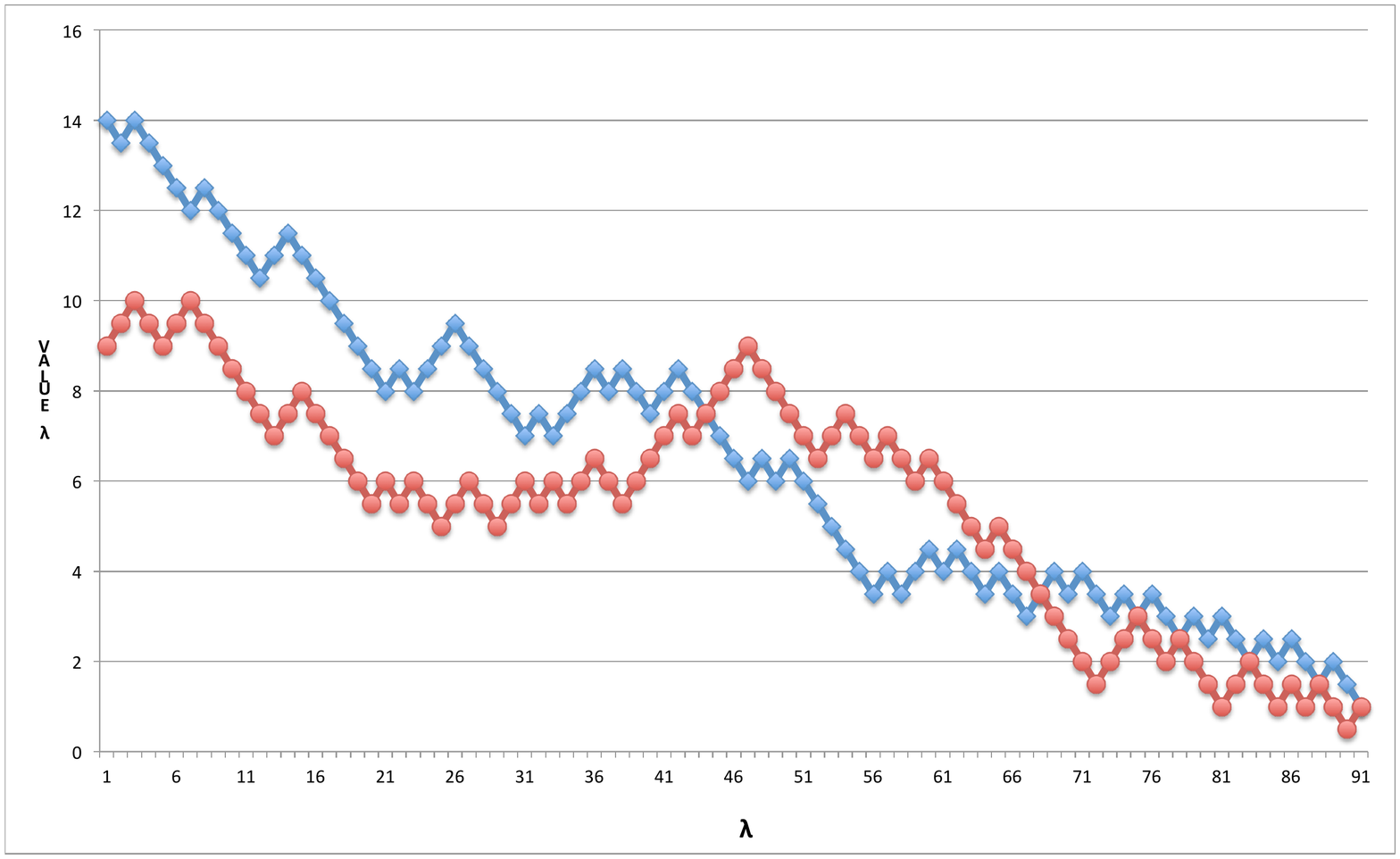}
\centering\caption{Values of labels $\lambda_{k,H}$, blue diamonds, and $\lambda_{k,V}$, orange  circles, ($k=1,\ldots,91$) for the $\beta-globin$ gene for Gallus. The value of  $\lambda_{0,H} = 12 $ and $\lambda_{0,V} = -6$  are not reported in the figure.}
 \label{HVbetaG}
 \includegraphics[width=15cm]{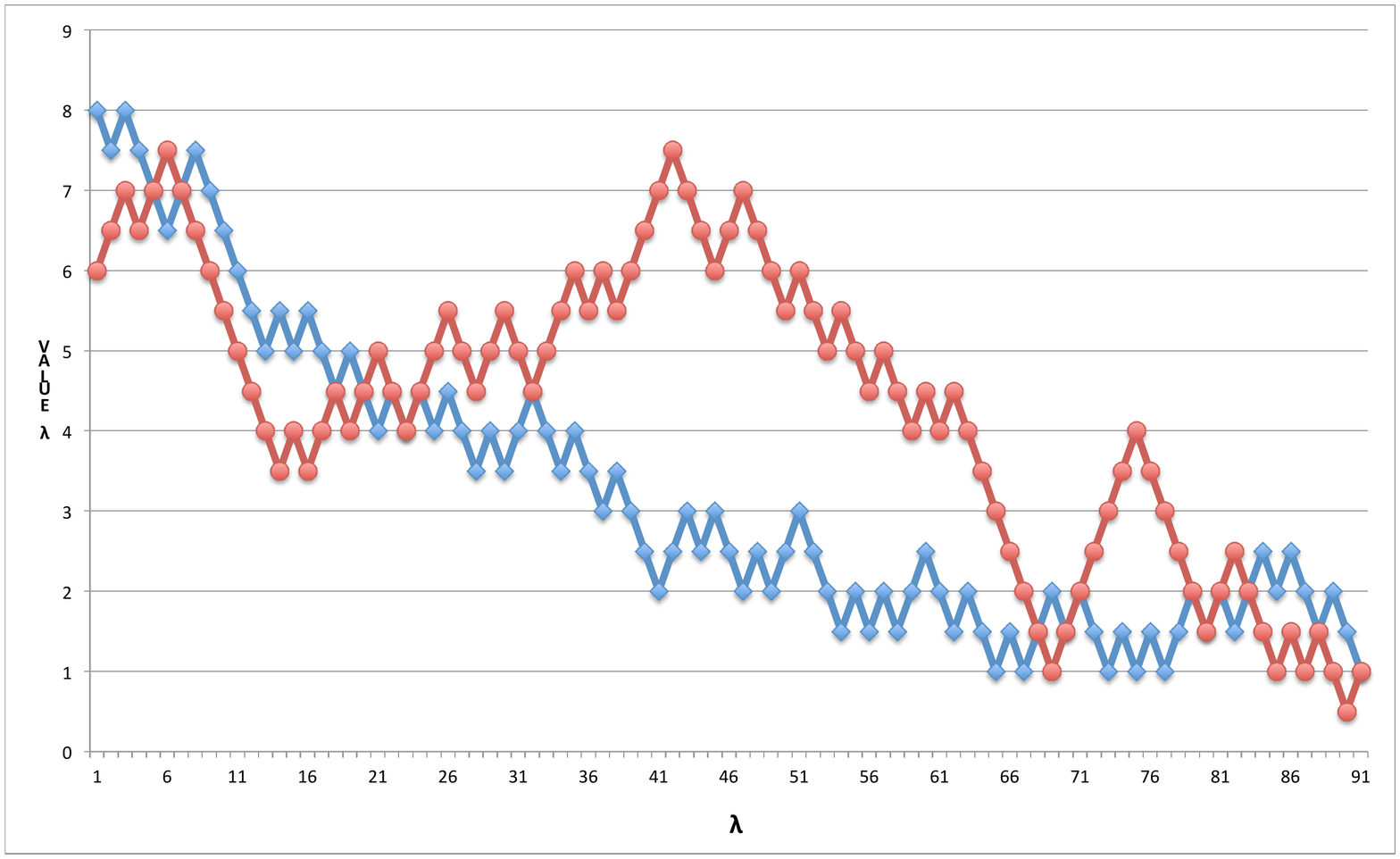}
\centering\caption{Values of labels $\lambda_{k,H}$, blue diamonds, and $\lambda_{k,V}$, orange  circles, ($k=1,\ldots,91$) for the $\beta-globin$ gene for  Opossum. The value of  $\lambda_{0,H} =  3$ and $\lambda_{0,V} = -4 $  are not reported in the figure.}
 \label{HVbetaO}
 \end{figure}

 Let us apply the above described algorithm to two more simple, but may not irrelevant, cases.
\begin{enumerate}
\item let us consider  the $N = 10$ sequence $ATGGTGCACC$ which corresponds to the first 10 nucleotides of the exon 1 in Human $\beta-globin$ gene, see below.\footnote{We use this sequence as it has been used in many examples in \cite{NHB} in order to allow an immediate comparison between our and other descriptions existing in the literature.} The sequence is labelled, the order  of the labels being specified in eq.(\ref{eq:Nnucl}),  by: 
\be
\mid ATGGTGCACC \rangle = \mid1,0;3,2; \five, \third;2,1; \five, \third; 2,1; \third,\third; 2,1;\third,\half;1,1 \rangle
\label{eq:10seq}
\ee
We report in  Fig. \ref{HVlabels} the diagram of the labels of the above string.
 In the  following, to simplify the notation, we denote
 \be
 \lambda_{0, a} \equiv J_{3, a} \;\;\;\;\;\;  \lambda_{k, a} \equiv J_{\alpha}^{N-k+1}  \;\;\;\;\;\;  k = 1,2,\dots,N-1 \;\;\;\;\;\;  a = H, V
 \ee

\item  let us consider  the $N = 92$ coding sequence of  the human
$\beta-globin$ gene, see Table 2 of \cite{Qi2012} (http://www.hindawi.com/journals/tswj/2012/104269/tab2/)
\bea
&& ATGGTGCACCTGACTCCTGAGGAGAAGTCTGCCGTTACTGCCCTGT\\ \nonumber
&&  GGGGCAAGGTGAACGTGGATGAAGTTGGTGGTGAGGCCCTGGGCAG
\eea
We report in  Fig. \ref{HVbetaH} the labels of the above sequence. In Figg. \ref{HVbetaG}-\ref{HVbetaO} the labels for the $N = 92$  sequence of  the  
$\beta-globin$ for Gallus and Opossum are reported, the nucleotide sequences being omitted to save space, see Table 2 of \cite{Qi2012}.
\end{enumerate}

Once  a discrete function is associated to a DNA/RNA sequence, one may study this function from many points of view. Let us cite just a few:  one can
\begin{itemize}
\item perform discrete circular convolution looking for symmetry, e.g. for function even or odd under $z$ ($z \in {\bf Z} $) circular symmetry, 

\item look for other kinds of symmetry, in the whole sequence or in some subsequence, for example:

\begin{enumerate}
\item In Fig. \ref{HVlabels}, neglecting the point n = 0, and subtracting 2 to the values of the H labels, one remarks that the function shows an inverse specular symmetry, i.e. is odd, around the point n = 5, that is $\lambda_{mp - n,H} =  -\lambda_{mp + n,H}$, where $\lambda_{mp,H}$ is the value of the  H label in the middle position ($mp = 5$). The meaning of this symmetry is that the number of not \emph{contracted} S and W nucleotides   between the positions $N +1 -mp + n = 6 + n$ and  $N +1 -mp - n = 6 -n$  is given by  $\lambda_{mp - n,H} - \lambda_{mp + n,H}$. Of course this information  can be obtained out from an analysis of the labels $\lambda_{k,H}$, but  the graphical representation may help to more quickly extract the information;
 
\item In  Fig. \ref{HVbetaH}, (human $\beta-globin$) one can also notice that: in the case of V labels, considering  approximately the range  of labels from 7 to 70  and subtracting 5 to the values of $\lambda_{k,V}$ and  in the case of H labels, considering  approximately the range  of labels from 3 to 50  and subtracting 7 to the values of $\lambda_{k,H} $, an approximate specular symmetry does show around the middle point of each series of labels;

\item In  Fig. \ref{HVbetaG}, ($\beta-globin$ of Gallus) the 9 labels on the right of the label $\lambda_{28,V}$ have the same value of  the 9 labels on the left.  The meaning of this symmetry  can be understood on the light of the above  remark;

\item In  Fig. \ref{HVbetaO}, ($\beta-globin$ of Opossum) the 8 labels on the right of the label $\lambda_{73,H}$ have the opposite value of  the 8 labels on the left subtracting to all labels the value 2,5;

\item In  Fig. \ref{HVbetaO},  the first 45 positive H labels fit a simple  linear  regression with a factor $R^Ó \approx 0,9$. This means that first 45 nucleotides are almost or purine or yrimidine. Indeed a linear segment in a graphic of the H (V) labels means that the nucleotides are all S or W  nucleotides (resp. Y or R).
\end{enumerate}

\item  study  the behavior under shifts of a sequences of length $M < N$;

\item analyze the  ``time reversal'' of the function, that is the function obtained by reading the DNA frame in the inverse direction $3 \prime \ra 5\primeÕ$ direction;

\item sample the sequence with windows of variable length, suggested by preliminary analysis of the discrete function or by the interest to focus on some particular sub-sequences.
\end{itemize}

Let us remark that from Fig.  \ref{HVbetaH}  a some kind of correlation does show between the H and V labels, the correlation factor between the two sets of labels  (not including $J_{3,H}$ and   $J_{3,V}$) being $r \approx 0,73$.  However, at the present stage of our analysis, we cannot state the this weak correlation is significant. Indeed the correlation between the $J_H$ and   $J_V$ labels  for the $ \beta-globin$ sequence  for Gallus  is $r \approx 0,81$ and for Opossum  $r \approx 0,53$.

Let us emphasize once more that the knowledge of all the labels $\; J_{3,H}, J_{3,V}; J_{H} ^N, J_{V} ^N; \ldots ;J_{H} ^2, J_{V} ^2 \;$ allows to  reconstruct the sequence in terms of the four letter nucleotide alphabet and conversely.
 
\section{Metrics and harmonic analysis}

\hspace{0.5 cm} Once established a correspondence between any DNA/RNA sequence and the value of a discrete function in ${\bf Z}$, according to the procedure  illustrated in the previous Section, it is possible to go further, in particular to define a {\bf distance} between two sequences and to perform the {\bf Discrete Fourier Transform} (DFT) of a sequence. 

\subsection{Distance between two sequences of equal length} 

\hspace{0.5 cm} The most simple definition of distance, based on two parameters $\alpha_i$ is the following one  (1-norm): given two $N$-elements strings the distance is defined by
\be
d^1(1,2) = \sum_{i= H,V} \; \alpha_i \; \left(  \sum_{k = 0}^{N-1} \; |\lambda^1_{k,i} - \lambda^2_{k,i} | \right)
\label{eq:dist}
\ee
where  1,2 denotes the two sequences and  $\alpha_i$ is a real parameter denoting the  ``scale''  of the distance in the  $i-th$-direction. Rigorously speaking eq.(\ref{eq:dist}) is the  1-norm metrics on the function 
\be
\Psi_N \equiv   \sum_{i= H,V} \; \alpha_i \; \left(  \sum_{k = 0}^{N-1} \, \lambda_{k,i} \right)
\label{eq:1norm}
\ee
computed for the two sequences. Eq.(\ref{eq:1norm}) is the natural distance in the  ``Taxicab geometry'' and is also called the ``Manhattan''   or  ``City block distance'' , up tp the parameters $ \alpha_i$.

Let us remark that this distance can be seen as the minimum number of different states  between the states corresponding to the first and second sequence, assuming the change of the parameters $\lambda_{k,i}$ by step of one H or V spin flip.

Of course one may also introduce an Euclidean distance
 \be
d^E(1,2) = \sum_{i = H,V} \; \alpha_i \; \sqrt{\sum_{k = 0}^{N-1} \;  \left[  \left(\lambda^1_{k,i}- \lambda^2_{k,i}\right)^2\right]}
\label{eq:2norm}
\ee
which can be read as  the 2-norm metrics on the function eq.(\ref{eq:1norm}). However 

One may consider $\lambda_H$ and   $\lambda_V$  as respectively the real an imaginary part  of a complex number so establishing a correspondence between a nucleotide sequence and a set of N complex numbers
\be
\lambda_k^c = \lambda_{k,H} + i \lambda_{k,V}
\label{eq.com}
\ee
With this generalization it appears more natural to define a distance as the modulus of the complex number, that is the Euclidean distance eq.(\ref{eq:2norm}). However one should give some arguments to motivate the definition eq.(\ref{eq.com}),  for example an interpretation of the complex conjugate $\lambda_k^{c*}$.

As we are speaking of distance between states, one may wonder if there is a relation with the distance between Hilbert states in quantum mechanics.
With the  interpretation above sketched, the measure  eq.(\ref{eq:1norm})
is, in some sense, a statistical distance between the sequences, close to the distance between quantum states defined in \cite{Woot},  
\be
d_{stat}(1,2)  =  \sum_{p_1}^{p_2}\; \frac{p_x}{\sqrt{X}}
\ee
where $p_z$ is the probability for a  state $x$  between the states 1 and 2 and  $X$ is the total number of different intermediate states. In our case,   $X$ should be read as the number of different sequences  in the path from the first to the second sequence, going step by step, each step corresponding to a transition or to a transversion. In our case  $p_x$  is equal $ \forall \; x$.

Let us discuss, with a few simple examples, the difference between the Hamming ($d_H$) and Kimura ($d_K$)  measure, respectively with parameter $a$, parameter $\alpha$ (transition) and $
\beta$ (transversion), and the above introduced one eq.(\ref{eq:1norm}), which we call crystal basis measure $d_{CB}$
\begin{enumerate}
\item This is a very simple didactic case. Let us compute the distance between the following 4-strings $1 \equiv CAUC, 2 \equiv CUCG, 3 \equiv AUGU, 4 \equiv CUGA,
 5 \equiv CUGU, 6 \equiv CAGU$  we have
\bea
&& d_H(1,2) = 3a   \;\;\;\; d_H(3,4) = 2 a   \;\;\;\;  d_H(5,6) =  a\nonumber \\
&& d_K(1,2) =   \alpha + 2\beta \;\;\;\; d_K(3,4) = 2 \beta \;\;\;\; d_K(5,6) =  \beta   \nonumber \\
&& d_{CB}(1,2)  = \alpha_H + 2 \alpha_V \;\;\;\; d_{CB}(3,4)  = 4 \alpha_H +  \alpha_V \;\;\;\; d_{CB}(5,6)  =   \alpha_V 
\label{eq:4stringd}
\eea
where the sequences are labelled as:
\bea
 CAUC =  \mid0,1;1,1; \half, \half; 0,0 \rangle  \;\;\;\; &  \;\;\;\;  CUCG =  \mid1,1;1,1; \half, \third; 0,1 \rangle \nonumber \\
 AUGU =  \mid-1,0;1,1; \third, \half; 1,1 \rangle  \;\;\;\; &  \;\;\;\;  CUGA =  \mid0,0; 0,1; \half, \third; 0,1 \rangle \nonumber \\
 CUGU =  \mid0,1;0,1; \half, \half; 0,1 \rangle  \;\;\;\; &  \;\;\;\;  CAGU =  \mid0,0; 0,1; \half, \half; 0,0 \rangle 
 \eea
 The above defined distances  are very simple, but they show the drawbacks to be explicitly dependent from somehow arbitrary parameters and implicitly from the length of the sequence. It is convenient do define normalized distance in the following way:
 
 \begin{enumerate}
 
 \item  the standard Hamming distance, $d^{st}_H$, the percentage of different symbols, that is the number of different symbols divided by the sequence length
 \be
 d^{st}_H \equiv \frac {d_H}{a N}
 \ee

\item the standard 2-parameters Kimutra distance $d^{st}_K$ \cite{Kimura80}
\be
d^{st}_K  \equiv -  \frac{1}{2} \,  \ln \, (( 1 -  2p - q ) \, \sqrt{1 - 2q}) =  -  \frac{1}{2} \,  \ln \, ( 1 -  2p - q ) -  \frac{1}{4} \,  \ln \, ( 1 -  2q )
\ee
where $p$ ($q$) is the percentage of positions  which are transitionally (resp. transversionally)  different, that is ($n_{tns}$ and $n_{tnv}$ denoting resp. the number of transitions and transversion)
\be
p \equiv \frac {n_{tns}}{N}  \;\;\;\;\;\;\;\;\; q \equiv \frac {n_{tnv}}{N} 
\ee

\item we can define the normalized crystal basis distance as it follows
\be
d^n_{CB}(1,2)\equiv  \sum_{i= H,V} \; \left(  \sum_{k = 0}^{N-1} \; \frac{ |\lambda^1_{k,i} - \lambda^2_{k,i} |}{|\lambda^1_{k,i}| + |\lambda^2_{k,i}|} \right)
\label{eq:distn}
\ee
Eq.(\ref{eq:distn}) is known in the literature as the   ``Canberra distance''.  Note the  eq.(\ref{eq:distn})  is not rigorously a distance as it does not satisfy the triangle assiom. In order to get a normalized rigorous distance one should use the normalized form of the Euclidean distance
 eq.(\ref{eq:2norm}).

 \end{enumerate}

Eq.(\ref{eq:4stringd}), using the above defined measures, becomes
 \bea
&& d^{st}_H(1,2) = 3/4   \;\;\;\; d^{st}_H(3,4) = 1/2   \;\;\;\;  d^{st}_H(5,6) =  1/4 \nonumber \\
&& d^{st}_K(1,2),  \;\; d^{st}_K(3,4)  \;\; \mbox{undefined } \;\;\;\;  d^{st}_K(5,6) =  - 1/2 \,  \ln 3/4  -  1/4 \,  \ln    1/2  \approx 0,32 \nonumber \\
&& d^n_{CB}(1,2)  = 5/2  \;\;\;\; d^n_{CB}(3,4)  = 4   \;\;\;\; d^n_{CB}(5,6)  =   1
\eea
 \item Let us consider the 6 sequences of Table \ref{table-1} as obtained  from the ancestral sequence $\, CCCCC$, by two point mutations  $C \ra U$. The Hamming and Kimura distance from the ancestral sequence is, for all the 6 sequences,
 \be
 d_H = 2a \;\;\;\; \;\;\;\;  d_K = 2 \alpha \;\;\;\; \;\;\;\; d^{st}_H = 2/5  \;\;\;\; \;\;\;\;  d^{st}_K =  - 0,5 \ln (1- 4/5) \approx 0,80
 \ee
The crystal basis measure  between the ancestral sequence $d_{CB,i}$ (labelled by $(k = 0, \ldots, 4) \, \lambda_{k,H}\equiv \mid \five; \five; 2; \third; 1\rangle$) and the i-th sequence (the sequences being ordered as in Table \ref{table-1}) is  almost different for any sequence and can be computed as 
\bea
&&  d_{CB,1} = 2  \alpha_H \;\;\;\; d_{CB,2} = 6  \alpha_H \;\;\;\;d_{CB,3} = 6  \alpha_H \nonumber \\
&& d_{CB,4} = 5  \alpha_H \;\;\;\;d_{CB,5} = 7  \alpha_H \;\;\;\;d_{CB,6} = 8  \alpha_H \nonumber \\
&&  d^n_{CB,1} = 2/3  \;\;\;\;\; d^n_{CB,2} = 11/4    \;\;\;\; d^n_{CB,3} = 13/6  \nonumber \\
&& d^n_{CB,4} = 5/3 \;\;\; \; \; d^n_{CB,5} = 17/6    \;\;\;\;d^n_{CB,6} = 23/6   
\label{eq:as}
 \eea
 From eq.(\ref{eq:as}) we read that the most distant sequence from the ancestral one is the one where the mutations   $C \ra U$ have occurred in the 2nd and 4th position while the most close is the one where the nucleotides have mutated in the first two positions. This example shows clearly that the distance $d_{CB}$ is not only depending from the number of the substitutions in the nucleotides in the sequences, but from the positions where the substitutions have occurred, a peculiar feature which may be of biological interest.

 \item Let us compute the distance  between the $N = 92$ coding sequence of  
$\beta-globin$ gene  for Homo sapiens, Didelphis virginiana  (Opossum) and Gallus gallus (see web site 

http://www.hindawi.com/journals/tswj/2012/104269/tab2/ for the nucleotide sequences and  Figg. \ref{HVbetaG}-\ref{HVbetaO} for the values of the $J_H$ and $J_V$).

  \begin{table}[htbp]
\begin{center}
\begin{tabular}{|c||c|c|c|}
\hline
  & $d_H$ & $d_K$ &    $d_{CB}$     \\
 \hline\hline
Homo/Gallus &  24 a & $10 \alpha + 14 \beta$   &  $105 \alpha_H + 129 \alpha_V$  \\
\hline\hline
Homo/Opossum &   24 a & $11 \alpha + 13 \beta$   &  $214 \alpha_H + 78 \alpha_V$  \\
 \hline\hline
Gallus/Opossum &   28 a & $6 \alpha + 22 \beta$   &  $313 \alpha_H + 125 \alpha_V$  \\
 \hline\hline
  & $d^{st}_H$ & $d^{st}_K$ &    $d^{n}_{CB}$     \\
 \hline\hline
Homo/Gallus &   $\approx 0,26 $  & $\approx 0,32 $   &  $\approx$ 20,25 \\
\hline\hline
Homo/Opossum &  $\approx 0,26 $   & $  \approx 0,26 $     &  $\approx$ 33,43\\
 \hline\hline
Gallus/Opossum & $\approx 0,30 $   &  $  \approx  0,10$    & $\approx$ 43,20  \\
 \hline
\end{tabular}
\caption{Distance between  the 92 nucleotides sequence of  the $\beta-globin$ of Homo, Gallus and Opossum. In the first column the considered couple of species.}
\label{table-disbetan}
\end{center}
\end{table}
 
From Table \ref{table-disbetan} we remark that
 the most distant (the most close)  $\beta-globin$ sequences, according to our parametrization, are those between Gallus and Opossum  (resp. between Homo and Gallus) in agreement with  estimates of \cite{Yu2011} and, partially, with those of \cite{Tod2008}.  Differently the values of the Hamming and Kimura distance are not at all reliable.

Let us point out once more that the value of  $d_H(1,2)$ ($d_K(1,2)$) depends only on the number of different nucleotides (resp. on the number of  transitions and  transversions) between the sequence 1 and 2, not from their position, while $d_{CB}(H,G)$ depends from  the number of different nucleotides, from the nature of the nucleotides as well as from their position; 
$n_1$ transitions and $n_2$ transversions in different positions will in general  lead to a different value of $d_{CB}$. Indeed a mutation of nucleotide in the m-th position changes the values of
$J^m_i$ and of $J_{3,i}$ and may change the value of  $J^k_i$, \;$k > m$. 
\end{enumerate}

Remarks:
\begin{itemize}

\item The  $d_{CB}$ distance is naturally defined for sequences of the same length. It may be extended to a distance between two sequences of different length by some suitable scaling procedure, but care must be paid for preserving the nature of integer or half-integer numbers of the labels $ \lambda_i$.

\item As $\lambda_{0,i} = J_{3,i}$  labels  different states in the same irrep.,  one may introduce a different parameter for this label, that is to replace  eq.(\ref{eq:1norm}) by
\be
\hat{\Psi} _N \equiv \sum_{i= H,V} \; \alpha_i^0 \; \left(\lambda_{0,i} \right) + \sum_{i= H,V} \; \alpha_i \; \left(  \sum_{k = 1}^{N-1} \, \lambda_{k,i} \right)
\label{eq:1norm-1}
\ee
However this choice is irrelevant if one makes use of the normalized distance.

\end{itemize}

 \subsection{Discrete Fourier Transform of a sequence}
 
\begin{figure}[htbp]
  \includegraphics[width=15cm]{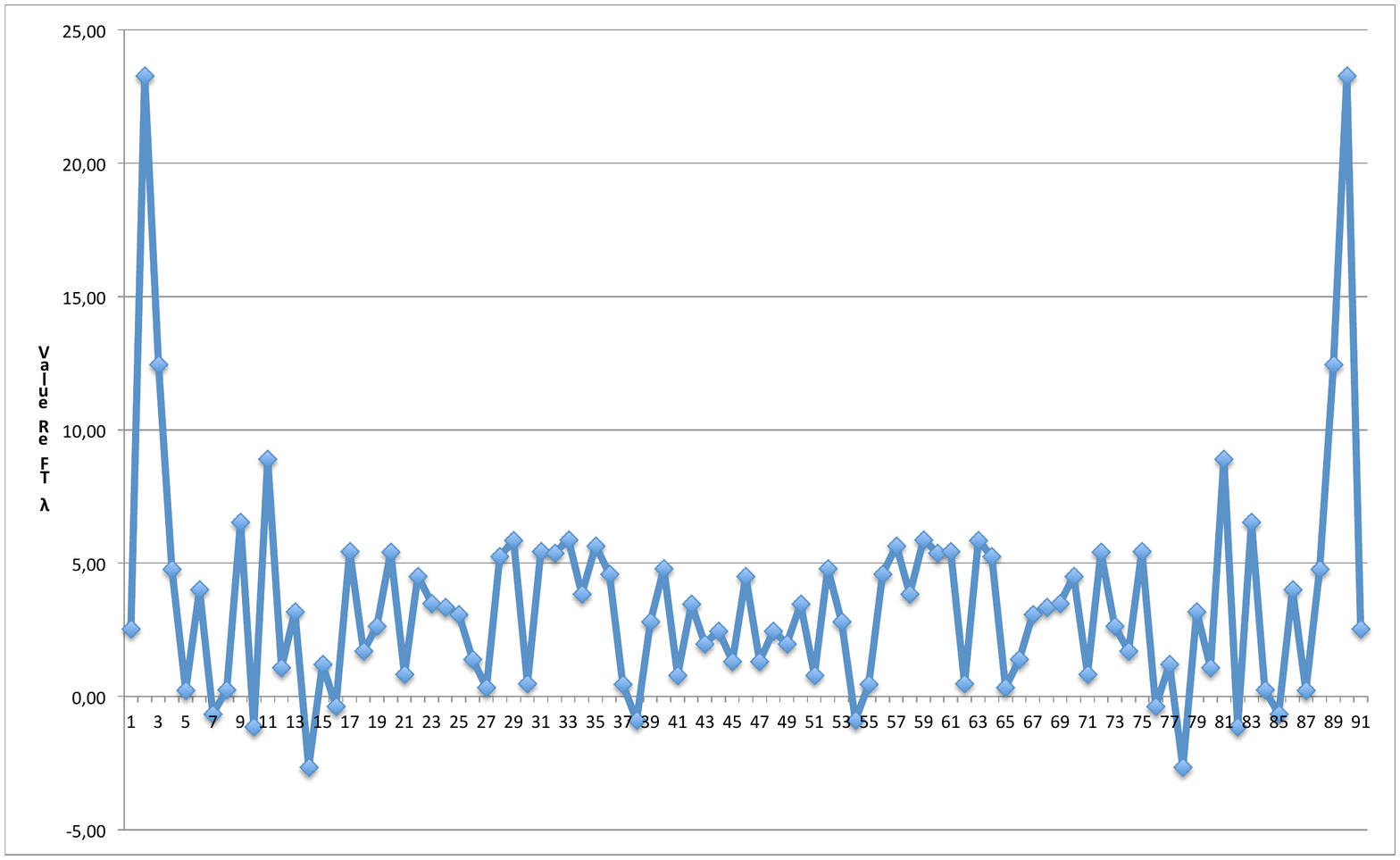}
\centering\caption{Values of the real part of the DFT of  $\lambda_{k,H}$ ($k=1,\ldots,91$) for the $\beta-globin$ gene for Homo sapiens.  The value of the real part of the DFT of  $\lambda_{0,V}$, not reported in the figure, is 512,5. }
 \label{ReFTBGH-H} 
 \includegraphics[width=15cm]{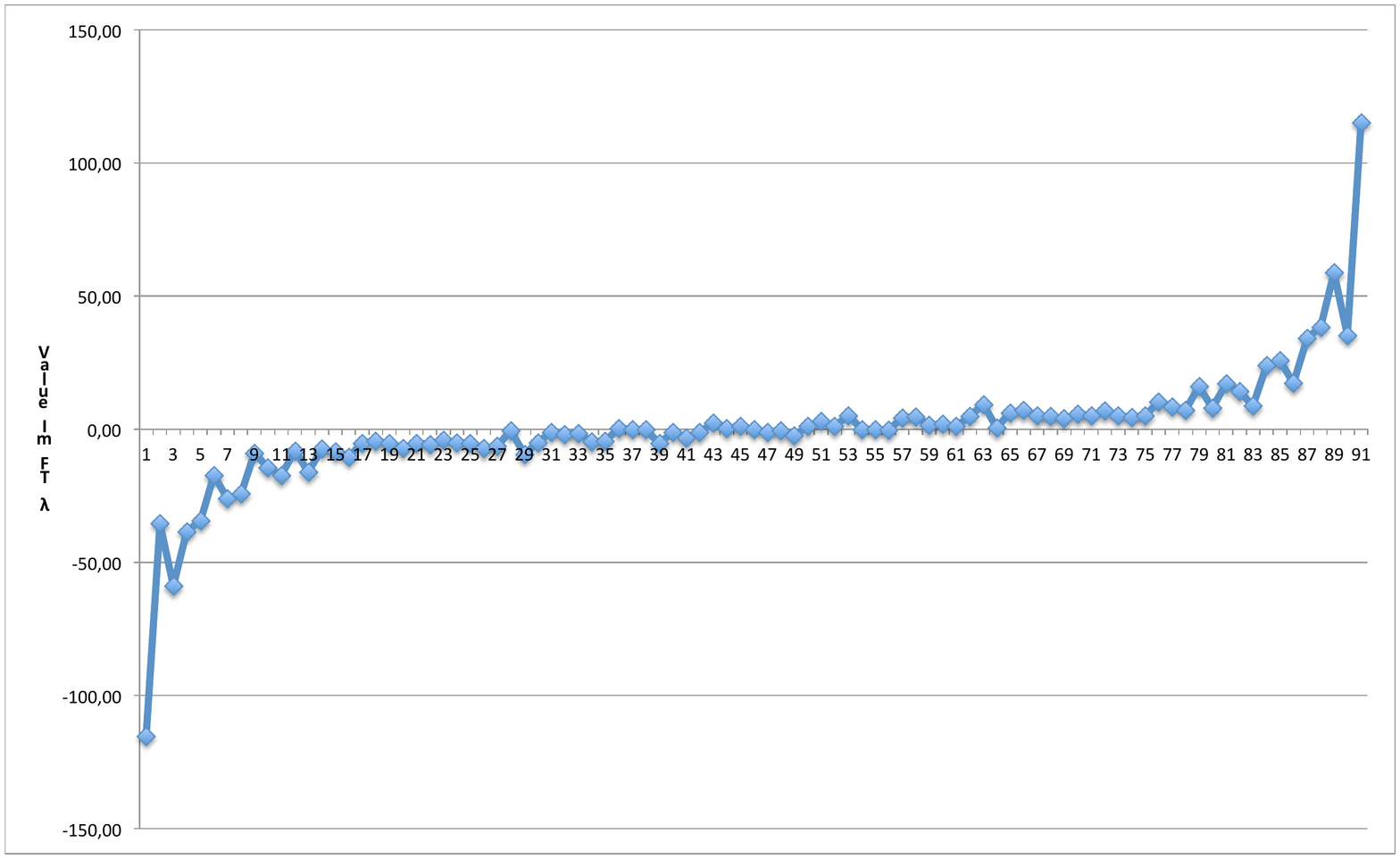}
\centering\caption{Values of the imaginary part of the DFT of  $\lambda_{k,H}$ ($k=1,\ldots,91$) for the $\beta-globin$ gene for Homo sapiens. The value of imaginary part of the DFT of  $\lambda_{0,V}$, not reported in the figure, is 0, as expected.}
 \label{ImFTBGH-H}
 \end{figure}

\begin{figure}[htbp]
  \includegraphics[width=15cm]{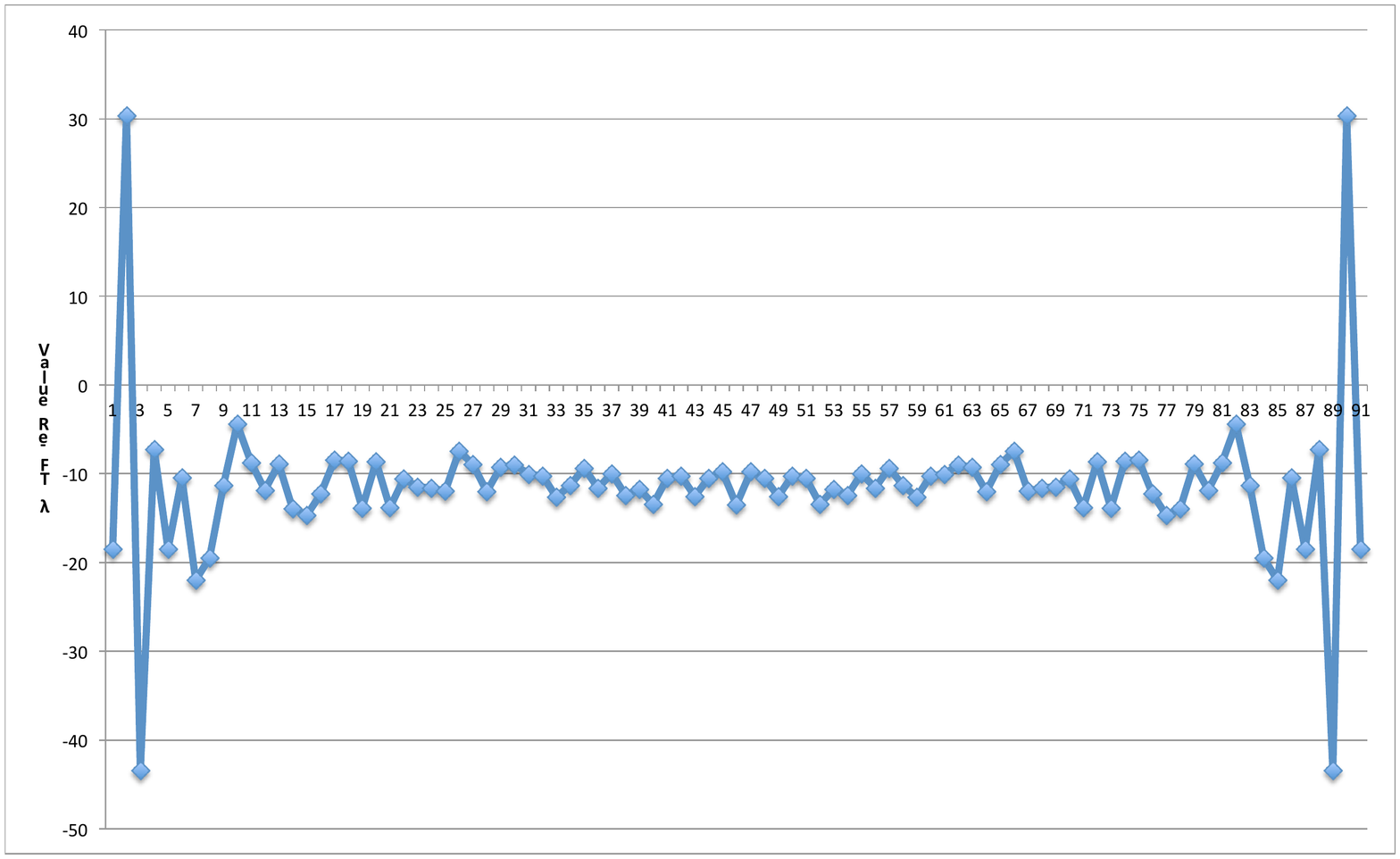}
\centering\caption{Values of the real part of the DFT of  $\lambda_{k,V}$ ($k=1,\ldots,91$) for the $\beta-globin$ gene for Homo sapiens. The value of the real part of the DFT of  $\lambda_{0,V}$, not reported in the figure, is 384,5. }
 \label{ReFTBGH-V} 
 \includegraphics[width=15cm]{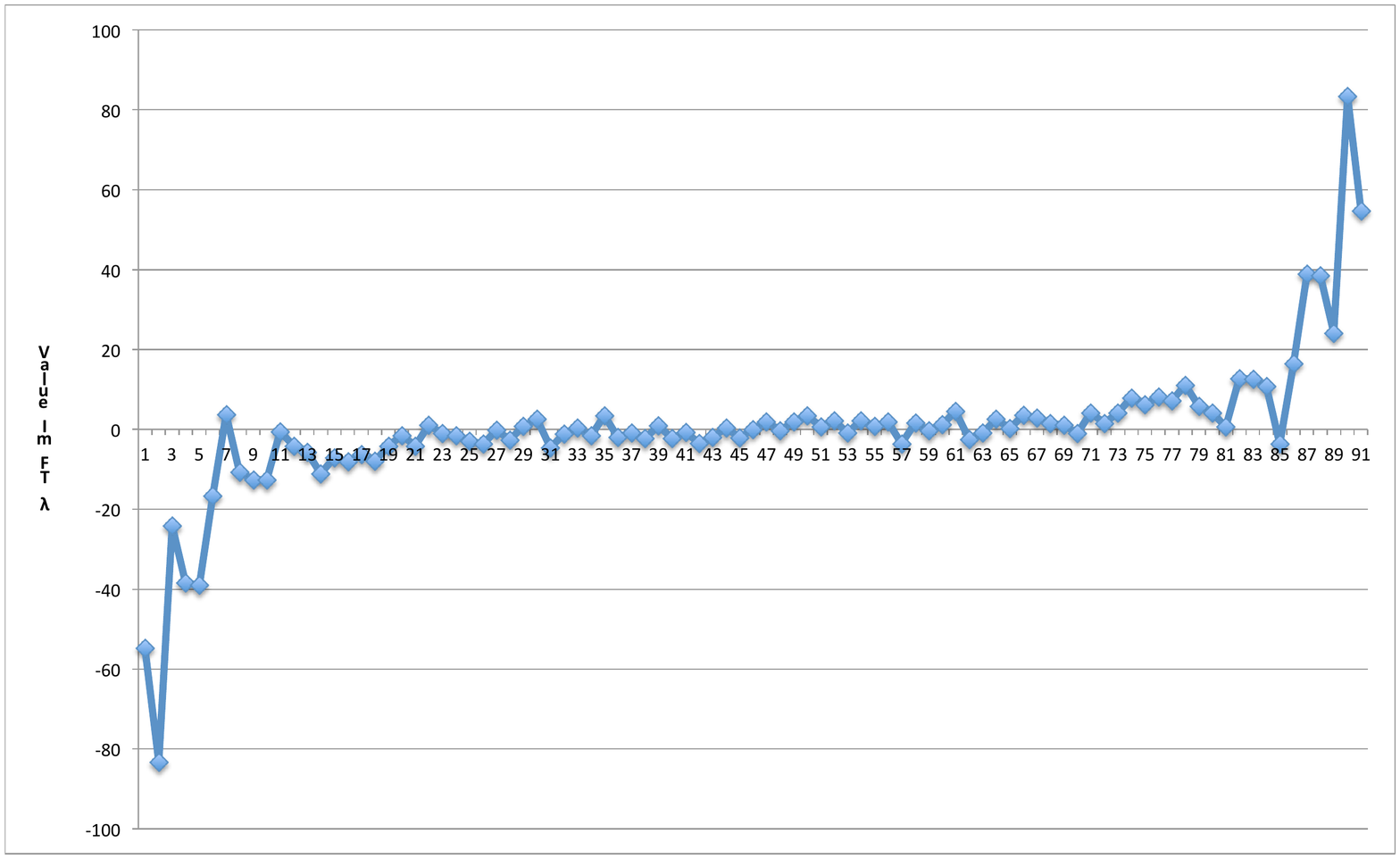}
\centering\caption{Values of the imaginary part of the DFT of  $\lambda_{k,V}$ ($k=1,\ldots,91$) for the $\beta-globin$ gene for Homo sapiens. The value of imaginary part of the DFT of  $\lambda_{0,V}$, not reported in the figure, is 0, as expected.}
 \label{ImFTBGH-V}
 \end{figure}

 \hspace{0.5 cm} The DFT of a numerical sequence is extremely useful because it allows to reveal periodicities in the input data as well as  the relative strength of the periodic components and is defined by\footnote{See Wikipedia, https://en.wikipedia.org/wiki/Discrete\_Fourier\_transform. }
\be
\wh{\lambda_n} = \sum_{k = 0}^{k = N-1} \;  \lambda_k \, e^{-\frac{2 i \pi k n}{N}}  \;\;\;\; \;\;\;\;  n = 0, 1, \ldots, N-1
\label{eq:DFT}
\ee
The inverse DFT is obtained as
\be
 \lambda_n = \frac{1}{N} \;  \sum_{k = 0}^{k = N-1} \;  \wh{\lambda_k} \, e^{\frac{2 i \pi k n}{N}}   
 \ee
 For real $\lambda_n$, which is our case, the above equation takes the  simpler form
\bea
 \lambda_n    & = &  \frac{1}{N}\; \sum_{k = 0}^{k = N-1} \;  \left[Re(\wh{\lambda_k}) \cos \frac{2  \pi k n}{N}
- Im(\wh{\lambda_k}) \sin \frac{2  \pi k n}{N}\right] \nonumber \\
& = & \frac{1}{N}\; \sum_{k = 0}^{k = N-1} \; \left[| \wh{\lambda_k} | \cos  \left(\frac{2  \pi k n}{N} + \Phi_k \right)\right]  
\label{eq:InDFT}
\eea
where
\be
| \wh{\lambda_k} | = \sqrt{Re^2(\wh{\lambda_k})+Im^2(\wh{\lambda_k})} \;\;\; \;\;\;   \;\;\; \Phi_k  = \arctan \left(\frac{Im(\wh{\lambda_k})}{Re(\wh{\lambda_k})} \right)
\ee
 Eq.(\ref{eq:InDFT}) can be rewritten,  N even, as
\be
 \lambda_n =  \frac{1}{N}\;\left[ \wh{\lambda_0} + 2 \sum_{k = 1}^{k = N/2-1} \, | \wh{\lambda_k} |  \cos  \left(\frac{2  \pi k n}{N} + \Phi_k \right)+   \wh{\lambda_{N/2}}  \cos   n \pi  \right]
\label{eq:IDFT1}
 \ee
 and, N odd,     as
\be
 \lambda_n =  \frac{1}{N}\;\left[ \wh{\lambda_0} + 2 \sum_{k = 1}^{k = (N+1)/2-1} \, | \wh{\lambda_k} |  \cos  \left(\frac{2  \pi k n}{N} + \Phi_k \right) \right]
\label{eq:IDFT2}
 \ee
 For real $\lambda_n$ it can be useful to rewrite eq.(\ref{eq:DFT}) in the following form, which helps to understand the features of the  Figg. \ref{ReFTBGH-H}-\ref{ImFTBGH-V} 
 \bea
 Re(\wh{\lambda_n})  &  = &  \sum_{k = 0}^{k = N-1} \;  \lambda_k \, \cos \frac{2  \pi k n}{N} \nonumber \\
  Im(\wh{\lambda_n})  &  = &  -\sum_{k = 0}^{k = N-1} \;  \lambda_k \, \sin \frac{2  \pi k n}{N}
 \label{eq:DFT-ri} 
  \eea
 The first term of eq.(\ref{eq:IDFT1})-(\ref{eq:IDFT2}) ($Re \,  \wh{\lambda_0} = \wh{\lambda_0} =  \sum_{k = 0}^{k = N-1}   \lambda_k $,  $Im \,  \wh{\lambda_0} =  0$) can be read as the average value of a constant  real component, the terms in the sum are sine waves of frequency k/N, amplitude  $2  | \wh{\lambda_k} |/N$ and phase shift $ \Phi_k$ and the last term is the  (real) coefficient of the highest frequency (= 1/2), not present for N odd.
 
  The challenge to define Fourier transforms for DNA sequences, more generally for strings of symbolic data, has already been faced in the literature, one of the main aim being the detection of periodicities in DNA or
protein sequences. The simplest solution is to map each symbol to a number, but, without an underlying algebraic framework, in general this choice is rather arbitrary and the result depends from the numerical labelling of the symbols, the nucleotides in the DNA/RNA case. Therefore several distinct and unrelated approaches exist. At our knowledge the first paper on this subject was published in 1986 \cite{Silverman86} assigning to each letter of the DNA alphabet a vertex of a regular tetrahedron.  Since then, many different DFT methods have been proposed and extensively used to study periodicities and repetitive elements in DNA sequences, genomes and protein structures.  Indeed a few years later in \cite{TaGi1989} a numerical sequence, called the indicator sequence, of numerical values $0$ and $1$ has been associated to a symbolic sequence and in \cite{Coward1997}  the two methods have been compared.  Other approaches have been proposed and 
in \cite{Vera2004a} the authors discuss  the equivalence and connections between several different and distinct approaches. In general the DFT are different for different approaches, even if  sometimes they share common features. Moreover  in \cite{Yin2014} an euclidean distance between the power spectra of the DFT of two sequences, even of different length, has been proposed and applied in particular for hierarchical clustering. However all these approaches are based on the assignment of a numerical indicator, different for any nucleotide, to the nucleotide in the k-th position. 
 
 In the present paper, making explicit use  of the underlying algebraic structure definite in the crystal basis model,  the DNA symbolic sequence is transformed in a numerical sequence, build up from the whole sequence composition, on which the DFT can be defined in the standard way.  Once introduced the DFT one may apply all the tools of this transform.  The DNA/RNA sequence is seen as a {\bf signal}, which can be decomposed, through the DFT, into different ``frequency'' components. Various filtering processing can be carried out as needed for the specific application, making more easily the manipulation of  the coefficients for different frequency components. When we try to determine the ``frequency content'',  we are trying to decompose a complicated signal into simpler parts. Many physical signals are best described as a sum of many individual frequencies components instead of the time domain description. The job of the DFT is  exactly to determine which frequencies a complicated signal is composed of. This translates for DNA/RNA signal to a description in terms  of the ``frequency content'' rather than a description in terms of the nucleotides content. This type of analysis may reveal a powerful tool, for example, in the analysis of genes, in the comparison of coding and not coding sequences  and for detecting periodicities. 
 
Of course the DFT  can  be directly performed on the the sequence of N nucleotides using the labels of eq.(\ref{eq:gc1}), but in this case the set of half-integer number  contains a limited amount of information, for example the Parseval theorem  ($\lambda_k \in \{ \pm 1/2\}$)
\be
\sum_{k = 0}^{k = N-1}|\lambda_k|^2 =  \frac{1}{N} \;  \sum_{n = 0}^{n = N-1} \;  |\wh{\lambda_n}|^2   \; \Lra \;  \sum_{k = 0}^{k = N-1} \, \left(\frac{1}{2} \right)^2  = 4^{-N}
\ee  
for any sequence of fixed length gives the same value, irrespective of the arrangement of the nucleotides in the sequence.

We report in Fig.\ref{ReTF} and Fig.\ref{ImTF}  the real, respectively the imaginary part, of the DFT of the labels of sequence eq.(\ref{eq:10seq}). Let us note that the symmetry, which shows in the two figures around the point 6, neglecting the point 0, is just a consequence of definition eq.(\ref{eq:DFT}), that is for $\lambda_k$ real the DFT shows the circular hermitian symmetry, i.e. $\wh{\lambda_{n,a}} = \wh{\lambda_{N-n,a}}^*$, where $^*$ denotes the complex conjugate.

 In Fig.\ref{ReFTBGH-H} and Fig.\ref{ImFTBGH-H} (Fig.\ref{ReFTBGH-V} and Fig.\ref{ImFTBGH-V})  we draw the real and the imaginary  part of the DFT of the H labels (resp. of V labels) of  human
$\beta-globin$ gene. We do not draw   $\wh{\lambda_{0,H}}  =  512,5$ and $\wh{\lambda_{0,V}}  =  384,5$.  In the figures we remark that, due to the circular hermitian symmetry,
the  real part   (the imaginary  part) of DFT coefficient  is symmetric (resp. anti-symmetric) with respect to the label 46. This property  reduces the analysis of the DFT coefficients of the labels of very long sequences of a one-half factor. Moreover we remark
 the presence of a peak  for $k = 2$  ($|\wh{\lambda_{2,H}} | \approx 23$, $|\wh{\lambda_{2,V}} | \approx 89$), corresponding to a frequency $k/N = 1/46$, and a secondary peak  for the case of V labels for $k = 1$ ($| \wh{\lambda_{1,V}} | \approx 57$),  corresponding to a frequency $k/N = 1/92$. Due to the circular symmetry these peak appear also for $k\prime = N - k$ with  shift of equal sign for H labels (of opposite sign for V labels). 
 
 The edge effects which appear in Figg. \ref{ReFTBGH-H}-\ref{ImFTBGH-V} are due to the fact that the  $\lambda_k$   in eq.(\ref{eq:DFT}) are real, with a specific pattern.
 
 The real part of the DFT of a motif of the same nucleotide ($NNN..N$,$ N \in \{C,G,U/T,A\}$), e.g. an isochore motif, is an horizontal segment, while the imaginary part is  an increasing straight segment.
 
\section{Conclusions}

\hspace{0.5 cm} We propose a mathematical modeling  in which a DNA/RNA sequence is represented by a numerical sequence, which somehow reflects the distribution of nucleotides in the original sequence.

  It is commonly believed that all the information for protein folding is contained in the coding RNA sequences. May the crystal basis parametrization of a sequence be useful for the investigation of this point ?
  
 From the proposed parametrization, it follows that a metrics can be defined between DNA/RNA sequences which may be useful to compute, e.g., which mutations  push away  more distantly a mutated sequence from the ancestral one.

The description allows, e.g.,  to look for symmetry of  subsequences from a completely new point of view and in Subsection 2.2 some  remarks and comments on this point are given. 

Moreover, as a sequence takes the form of a signal, it can be decomposed into its elementary components  performing the DFT of the signal. The  ``frequency content'' of a sequence should be better interpreted to look for a possible deeper biological  meaning than its mathematical definition. It seems that there is a dominant frequency $\nu = 2/N$, depending on the length of the sequence.

Let us emphasize once more that the aim of this paper is not to make specific applications of this mathematical description for DNA or RNA sequences, but to try to illustrate the method and to enlighten, without  any claim  of completeness, its potentiality and versatility.   Dealing with specific problems may possibly validate some of the possibilities here outlined.

The algebraic structure, underlying the model, illustrated in Subsection 2.1, which provides the framework for the mathematical parametrization, has been  exploited, in the present paper,  only  to represent a nucleotide sequence as a state in the N-fold tensor product of fundamental irreps..  The wealth of the algebraic structure can be further used to model a mutation or deletion or insertion of a nucleotide in a sequence as the action of suitable tensor operator.
Let us recall that  tensor operators  $\tau_{m}^{j}$ of $\mathcal{U}_{q \to 0}(sl(2))$, with the consequent extension of the Wigner-Eckart theorem,  have been defined in \cite{MS}. An explanation of the organization of the genetic code in this framework has been given in \cite{S} and a preliminary application to deletion from 4-nucleotide (3-nucleotide) to 3-nucleotide (resp. 2-nucleotide) has been discussed  \cite{FSS2}.

 The ($q \to 0$)-Wigner-Eckart theorem has the peculiar feature that the selection rules, that is the transitions between states which are permitted or forbidden,  do not depend only on the rank of the tensor operator and 
on the initial state, but in a crucial way from the specific component of the tensor in consideration. 
 
 For example, the modeling of a pure negative trannsversion ($\Delta J_H = 0, \Delta J_V = -1$, i.e. $C \ra G$ or $T \ra A$) as the tensor product of the identity operator on the H spin and a vector operator on the V spin (${\bf 1}_H \otimes \tau_{-1}^{1_V}$) acts on the state of human  $\beta-globin$ as it follows (we indicate only the values of the labels $J_{3,i}, J^N_i, i =H, V$)
\be
|9, -6; 11, 7>  \;\;\; \Lra \;\;\; |9, -7; 11, 8> 
\ee
and, in the considered case, does not give any restriction on the mutation.
On the contrary  the modeling of a pure positive trannsversion ($\Delta J_H = 0, \Delta J_V = 1$, i.e. $G \ra C$ or $A \ra T$)  as the tensor product of the identity operator on the H spin and a vector operator on the V spin (${\bf 1}_H \otimes \tau_{1}^{1_V}$) acts on the state of human  $\beta-globin$ as it follows 
\be
|9, -6; 11, 7>  \;\;\; \Lra \;\;\; |9, -6; 11, 6> 
\ee
and it implies that the mutation can only occur in a nucleotide in the $n-th$ position, $n \leq 81$.
This is a relatively weak bound, further constraints arise by  requiring the tensor operator to behave in a fixed way for the irreps. labelled by $J^K_i$, $K < N$, that is considering a kind of   ``russian doll tensor operator''.
 
We are hopeful that the presently proposed analysis may prove to be  an useful tool for elucidating the topology of genetic regulatory networks, for putting into evidence  possible existence of symmetry, for  pointing out possible DNA  ``fundamental frequencies'', for analyzing  exon and intron regions, cis-regulatory regions and  so on. 

Let us repeat that the approach here presented has to be further
 developed and, above all, its usefulness has to be validated by application to specific problem.

 \vspace{0.5cm}

 {\bf Acknowledgments} It is a pleasure to thank Paul Sorba for useful discussions and suggestions.

 \bibliography{DFTbiblio}

\begin{thebibliography}{21}
\providecommand{\natexlab}[1]{#1}
\providecommand{\url}[1]{\texttt{#1}}
\expandafter\ifx\csname urlstyle\endcsname\relax
  \providecommand{\doi}[1]{doi: #1}\else
  \providecommand{\doi}{doi: \begingroup \urlstyle{rm}\Url}\fi

\bibitem[Afreixo et~al.(2004)Afreixo, Ferreira, and Santos]{Vera2004a}
Vera Afreixo, Paulo~J.S.G. Ferreira, and Dorabella Santos.
\newblock Fourier analysis of symbolic data: A brief review of nucleotide
  sequences.
\newblock \emph{Digital Signal Processing}, 14:\penalty0 525--530, 2004.

\bibitem[Coward(1997)]{Coward1997}
E.~Coward.
\newblock Equivalence of two fourier methods for biological sequences.
\newblock \emph{J. Math. Biol.}, 36:\penalty0 64--70, 1997.

\bibitem[Frappat et~al.(1998)Frappat, A.Sciarrino, and Sorba]{FSS1}
L.~Frappat, A.Sciarrino, and P.~Sorba.
\newblock A crystal base for the genetic code.
\newblock \emph{Phys.Lett. A}, 250:\penalty0 214--221, 1998.

\bibitem[Frappat et~al.(2001)Frappat, A.Sciarrino, and Sorba]{FSS2}
L.~Frappat, A.Sciarrino, and P.~Sorba.
\newblock Crystalizing the genetic code.
\newblock \emph{J.Biol.Phys.}, 27:\penalty0 1--34, 2001.

\bibitem[Hamming(1950)]{Hamming}
R.W. Hamming.
\newblock Error detecting and error correcting codes.
\newblock \emph{Bell System Technical Journal}, 29\penalty0 (2):\penalty0
  147--160, 1950.

\bibitem[Hamori and Ruskin(1983)]{HR}
E.~Hamori and J.~Ruskin.
\newblock H curves, a novel method of representation of nucleotide series
  especially suited for long dna sequences.
\newblock \emph{J.Biol.Chem.}, 258\penalty0 (2):\penalty0 1316--1327, 1983.

\bibitem[Hosangadi(2012)]{SH}
Sandeep Hosangadi.
\newblock Distance measure for sequences.
\newblock arXiv 1208.5713, 2012.

\bibitem[Kashiwara(1990)]{Kashi}
M.~Kashiwara.
\newblock Crystallizing the $q$-analogue of universal enveloping algebras.
\newblock \emph{Commun.Math.Phys.}, 133:\penalty0 791--797, 1990.

\bibitem[Kimura(1980)]{Kimura80}
M.~Kimura.
\newblock A simple method for estimating evolutionary rate of base
  substitutions through comparative studies of nucleotide sequences.
\newblock \emph{J.Mol.Evol.}, 16:\penalty0 111--120, 1980.

\bibitem[Levenshtein(1966)]{Leve65}
V.I. Levenshtein.
\newblock Binary codes capable of correcting deletions, insertions, and
  reversal.
\newblock \emph{Soviet Physics Doklady}, 10:\penalty0 707--710, 1966.

\bibitem[Marotta and Sciarrino(2000)]{MS}
V.~Marotta and A.~Sciarrino.
\newblock Tensor operator and wigner-eckart theorem for $u_{q \rightarrow
  0}(sl(2))$.
\newblock \emph{J.Math.Phys.}, 41:\penalty0 5735--5744, 2000.

\bibitem[Nandy et~al.(2006)Nandy, Harle, and Basak]{NHB}
A.~Nandy, M.~Harle, and S.C. Basak.
\newblock Mathematical descriptors of dna sequences: development and
  applications.
\newblock \emph{ARKIVOC}, pages 211--238, 2006.

\bibitem[Nielsen and al.(2016)]{MIT}
Alec~A.K. Nielsen and al.
\newblock Genetic circuit design automation.
\newblock \emph{Science}, 352\penalty0 ((6281) aac7341):\penalty0 1--10, 2016.

\bibitem[Qi and al(2012)]{Qi2012}
Xingqin Qi and al.
\newblock Numerical characterization of dna sequence based on dinucleotides.
\newblock \emph{The Scientific World Journal}, page 104269, 2012.

\bibitem[Sciarrino(2003)]{S}
A.~Sciarrino.
\newblock A mathematical model accounting for the organization in multiplets of
  the genetic code.
\newblock \emph{BioSystems}, 69:\penalty0 1--13, 2003.

\bibitem[Silverman and Linsker(1986)]{Silverman86}
B.D. Silverman and R.~Linsker.
\newblock A measure of dna periodicity.
\newblock \emph{J.Theor.Biol.}, 118:\penalty0 295--300, 1986.

\bibitem[S.Tavare and Giddings(1989)]{TaGi1989}
S.Tavare and B.W. Giddings.
\newblock Some statistical aspects of the primary structure of nucleotide
  sequences.
\newblock In M.~S. Waterman, editor, \emph{Mathematical Methods for DNA
  Sequences}, Boca Raton, Florida (USA), pages 117--131. CRC Press, 1989.

\bibitem[Todeschini and al.(2008)]{Tod2008}
R.~Todeschini and al.
\newblock A new similarity/ diversity measure for the characterization of dna
  sequences.
\newblock \emph{Croat. Chem. Acta}, 81:\penalty0 657--664, 2008.

\bibitem[Wootters(1981)]{Woot}
W.K. Wootters.
\newblock Statistical distance and hilbert space.
\newblock \emph{Phys.Rev. D}, 23\penalty0 (2):\penalty0 357--362, 1981.

\bibitem[Yin et~al.(2014)Yin, Chen, and Yau]{Yin2014}
Changchuan Yin, Ying Chen, and Stephen S-T. Yau.
\newblock A measure of dna sequenc esimilarity by fourier transform with
  applications on hierarchical clusterin.
\newblock \emph{J.Theor.Biol.}, 359:\penalty0 18--28, 2014.

\bibitem[Yu(2011)]{Yu2011}
Hong~Jie Yu.
\newblock Similarity analysis of dna barcodes sequences based on compressed
  feature vector.
\newblock In D.~Huang and al., editors, \emph{Bio-Inspired Computing and
  Applications}, Lecture Notes in Bioinformatics. 7th International Conference
  on Intelligent Computing, ICIC 2011, Springer, 2011.

\end{thebibliography}

 \newpage
   \newpage
   \newpage
  \newpage
 
\end{document}